\documentclass[11pt]{article}
\usepackage{natbib}
\usepackage{amssymb,amsmath,amsthm,latexsym,amsfonts,amscd,dsfont}
\usepackage{a4wide,graphicx,tikz}
\usetikzlibrary{arrows,shapes,shadows,fit,backgrounds}

\tikzstyle{trader} = [circle, draw, top color=white, bottom color=blue!30, draw=blue!50!black!100, drop shadow, minimum height=4em]
\tikzstyle{bank} = [rectangle, draw, top color=white, bottom color=red!20, draw=red!50!black!100, drop shadow, rounded corners, minimum height=3em, text width=4em, text centered]
\tikzstyle{market} = [rectangle, draw, top color=white, bottom color=green!20, draw=green!50!black!100, drop shadow, rounded corners, minimum height=3em, text width=4em, text centered]
\tikzstyle{background} = [rectangle,fill=gray!10, inner sep=0.2cm, rounded corners=5mm]
\tikzstyle{line} = [draw, latex'-latex']
\tikzstyle{from} = [draw, latex'-]
\tikzstyle{to} = [draw, -latex']

\newcommand{\url}[1]{{\tt \small #1}}

\newcommand{\ind}[1]{1_{\{#1\}}}
\newcommand{\Exx}{\mathbb{E}}
\newcommand{\Ex}[2]{\mathbb{E}_{#1}\!\left[\,#2\,\right]}
\newcommand{\ExT}[3]{\mathbb{E}_{#1}^{#2}\!\left[\,#3\,\right]}
\newcommand{\Qxx}{\mathbb{Q}}

\newcommand{\rec}{R}
\newcommand{\lgd}{\mbox{L{\tiny GD}}}

\newcommand{\fva}{\mbox{F{\tiny VA}}}

\begin{document}

\title{Funding Valuation Adjustment:\\a consistent framework including CVA, DVA, collateral,\\netting rules and re-hypothecation}
\author{
Andrea Pallavicini \\ {\small Financial Engineering} \\ {\small Banca IMI, Milan}
\and
Daniele Perini \\ {\small Financial Engineering} \\ {\small Mediobanca, Milan}
\and
Damiano Brigo \\ {\small Dept. of Mathematics} \\ {\small King's College, London}
}
\date{
First version October 10, 2011.\emph{\medskip \medskip\ }This version \today.
}

\maketitle

\begin{abstract}

In this paper we describe how to include funding and margining costs into a risk-neutral pricing framework for counterparty credit risk. We consider realistic settings and we include in our models the common market practices suggested by the ISDA documentation without assuming restrictive constraints on margining procedures and close-out netting rules. In particular, we allow for asymmetric collateral and funding rates, and exogenous liquidity policies and hedging strategies. Re-hypothecation liquidity risk and close-out amount evaluation issues are also covered.

We define a comprehensive pricing framework which allows us to derive earlier results on funding or counterparty risk. Some relevant examples illustrate the non trivial settings needed to derive known facts about discounting curves by starting from a general framework and without resorting to ad hoc hypotheses.

Our main result is a bilateral collateralized counterparty valuation adjusted pricing equation, which allows to price a deal while taking into account credit and debt valuation adjustments along with margining and funding costs in a coherent way. We find that the equation has a recursive form, making the introduction of an additive funding valuation adjustment difficult. Yet, we can cast the pricing equation into a set of iterative relationships which can be solved by means of standard least-square Monte Carlo techniques.

\end{abstract}

\medskip

\noindent\textbf{AMS Classification Codes}: 62H20, 91B70 \newline
\textbf{JEL Classification Codes}: G12, G13 \newline

\noindent \textbf{keywords}: Funding Cost, Cost of Funding, Bilateral Counterparty Risk, Credit Valuation Adjustment, Debt Valuation Adjustment, Collateral Modeling, Margining Cost, Close-Out, Re-hypothecation, Default Correlation.

\pagestyle{myheadings} \markboth{}{{\footnotesize  A. Pallavicini, D. Perini and D. Brigo. Consistent Cost of Funding}}

\newpage
\tableofcontents
\newpage

\section{Introduction}

Cost of funding has become a paramount topic in the industry. One just has to look at the number of presentations and streams at modeling conferences that deal with this topic to realize how much research effort is being put into it. And yet literature is still in its infancy and there is very little published material.

Funding costs are linked to collateral modeling, which in turn has a strong impact on credit and debit valuation adjustments (CVA and DVA). While there are several papers that try to deal with these effects separately, very few try to build a consistent framework where all such aspects can live together in a consistent way. Our aim in this paper is to build such a framework. 

When dealing with funding costs, one has to make a decision first on whether to take a single deal (bottom-up) view or a large-pool, or homogeneous, cost view.

The bottom-up approach can deal with funding costs that are deal specific. It is also an approach that distinguishes between funding and investing in terms of returns since in general different spreads will be applied when borrowing or lending. This is not the unique possibility, however, and typically treasury departments in banks work differently. One could assume an average cost of funding (borrowing) to be applied to all deals, and an average return for investing (lending). This would lead to two curves that would hold for all funding costs and invested amounts respectively, regardless of the specific deal, but depending on the sign of the exposures.

One can go further with the homogeneity assumption and assume that the cost of investing and the cost of funding match, so that spreads are not only the same across deals, but are equal to each other, implying a common funding (borrowing) and investing (lending) spread. In practice the spread would be set by the treasury at a common value for borrowing or lending, and this value would match what is expected to go on across all deals on average. This homogeneous average approach would look at a unique funding/lending spread for the bank treasury to be applied to all products traded by capital markets. The homogeneous approach is assumed for example in \cite{Fries2010} and \cite{Piterbarg2010}.

In this paper we stay as general as possible and therefore assume a micro view, but of course it is enough to collapse our variables to common values to obtain any of the large pool approaches.    

We also point out that the bottom-up {\it vs.} large-pool approaches view the treasury department in very different ways. In the bottom-up model the treasury takes a very active role in looking at funding costs and becomes an operations center. In the homogeneous model the treasury takes more the role of a central supporting department for the bank operations. The second view is prevailing at the moment, but we point out that it is more difficult to implement absence of arbitrage in that framework.  

\subsection{Past Literature on Funding and Collateral}

The fundamental impact of collateralization on default risk and on CVA and DVA has been analyzed in \cite{Cherubini} and more recently in \cite{BrigoCapponiPallaviciniPapatheodorou} and \cite{BrigoCapponiPallavicini}. The works \cite{BrigoCapponiPallaviciniPapatheodorou} and \cite{BrigoCapponiPallavicini} look at CVA and DVA gap risk under several collateralization strategies, with or without re-hypothecation, as a function of the margining frequency, with wrong-way risk and with possible instantaneous contagion. Minimum threshold amounts and minimum transfer amounts are also considered. We cite also \cite{Brigo2011} for a list of frequently asked questions on the subject.

The fundamental funding implications in presence of default risk have been considered in \cite{MoriniPrampolini2011}, see also \cite{Castagna2011}. These works focus on particularly simple products, such as zero coupon bonds or loans, in order to highlight some essential features of funding costs.

\cite{Fujii2010} analyzes implications of currency risk for collateral modeling. 

An initial analysis of the problem of replication of derivative transactions under collateralization but without default risk and in a purely classical Black and Scholes framework has been considered in \cite{Piterbarg2010}. An extension of it accounting for bilateral counterparty risk is provided in \cite{BurgardKjaer2011}.

The above references constitute an important start for the funding cost literature but do not have the level of generality needed to include all the above features in a consistent framework that can then be used to manage complex products. The only exceptions so far are \cite{Crepey2011}, who however does not allow for credit instruments in the basic portfolio.

Our approach to introducing a general framework takes into account our past research on bilateral counterparty risk, collateral, re-hypothecation and wrong way risk, across asset classes. We then add cost of funding consistently, completing the picture. 

\subsection{Credit and Debt Valuation Adjustment (CVA and DVA)}

Prior to 2007, counterparty credit risk was accounted for through unilateral CVA (or UCVA), see for example \cite{BrigoMasetti} for the general framework under netting, and \cite{BrigoPallavicini2007}, \cite{BrigoBakkar}, \cite{BrigoChourdakis}, \cite{BrigoMoriniTarenghi} for applications of this framework to different asset classes including interest rates, equity, credit and energy commodities, all embedding wrong-way risk.

The unilateral assumption, implying the omission of the DVA term, is justified, for example, when one of the two parties in the deal can be considered as being default-free, an assumption that was applied to many financial institutions prior to 2007. However, if both parties are defaultable this method is incorrect as the valuation is asymmetric between the two parties and breaks the basic accounting principle according to which an asset for one party is a liability for another. This inconsistency is remedied with the incorporation of the DVA, leading to bilateral CVA, where consistency imposes the presence of first to default indicators in the two CVA and DVA terms. The first-to-default clause appears in \cite{BieleckiRutkowski2002} and was made explicit in \cite{BrigoCapponi2010} in the case of an underlying CDS. It was considered in the case of interest-rate portfolios in \cite{BrigoPallaviciniPapatheodorou} and also appears in \cite{Gregory2009}. In this context, the paper \cite{BrigoCapponiPallaviciniPapatheodorou} extends the bilateral theory to collateralization and re-hypothecation and \cite{BrigoCapponiPallavicini} shows cases of extreme contagion where even continuous collateralization does not eliminate counterparty risk.

A simplified approach would be to consider just the difference of unilateral CVA (UCVA) and the corresponding unilateral DVA (UDVA), neglecting first to default risk. The combination BCVA = UCVA - UDVA would define a simplified but inconsistent version of  bilateral CVA. The error in this simplification is analyzed in \cite{BrigoBuescuMorini}.

In 2009 ISDA modified the wording of the close-out rule in standard CSA agreements, see \cite{ISDA2009}, allowing for the possibility to switch from a risk-free close-out rule to a replacement criterion that, after the first default, includes the UDVA of the surviving party into the recoverable amount. Depending on the type of close-out that is adopted and on the default dependency between the two parties in the deal, this may lead to different types of inconsistencies highlighted for example in \cite{BrigoMorini2011}, \cite{BrigoMorini2010Flux}.

\subsection{Counterparty Risk and Basel III}

The Basel III Accord prescribes that banks should compute unilateral UCVA by assuming independence of exposure and default. Wrong-way risk is included through one-size-fits-all multipliers. The advanced framework allows banks to compute the effect of wrong-way risk using own models, while the standardized approach accounts for the effect by means of a one-size-fits-all multiplier. Examples in \cite{BrigoPallavicini2007}, \cite{BrigoBakkar}, \cite{BrigoChourdakis}, \cite{BrigoMoriniTarenghi} indicate that the actual multiplier is quite sensitive to model calibration and market conditions, and a hypothetical advanced framework might be more prudent. Interestingly, the Basel III Accord chooses to ignore the DVA in the calculation for capital adequacy requirements, although consideration of the DVA needs to be included according to accounting standards.

\subsection{Inclusion of Funding Costs (FVA)}

When including funding costs one has to make a number of choices. We already pointed out above that the first choice is whether to take a bottom-up view at single deal level or a large pool view. We adopt the former because more general, but most of the initial literature on funding takes the latter view, in line with current operational guidelines for treasury departments. Clearly the latter view is a particular case of the former, so that we are actually dealing with both views.

When we try and include cost of funding in the valuation of a deal we face a difficult situation. The deal future cash flows will depend on the funding choices that will be done in the future, and pricing those cash flows today involves modeling the future funding decisions. The dependence in not additively decomposable in the same way as credit valuation and debit valuation adjustments are in the case with no funding costs.

This leads to a recursive valuation equation that is quite difficult to implement, especially when dealing with products that are path dependent, since one needs at the same time backward induction and forward simulation. 

In this sense it is too much to expect that funding costs can be accounted for by a simple Funding Valuation Adjustment (FVA) term. Such term can be defined formally but would not add up with CVA and DVA terms in a simple way.

The recursion has been found also with different approaches, see for example \cite{Crepey2011} and \cite{BurgardKjaer2011}.

\subsection{Structure of the Paper}

In the following sections we calculate the price of a deal inclusive of counterparty credit risk (CVA and DVA), margining costs, and funding and investing costs. We start in section \ref{sec:BCCFVA} by listing the main findings. Then, we describe how to obtain them. In section \ref{sec:ISDA} we extend the results of \cite{BrigoCapponiPallaviciniPapatheodorou} to include margining costs when pricing a deal with counterparty credit risk, so that we are able to fix all the terms of pricing equation, but for funding and investing costs. In section \ref{sec:funding} we analyze the relationships between funding, hedging and collateralization to fix the remaining terms in the pricing equation.

Within the derivation of the results, we make some diversions to specialize the pricing formulae to simple but relevant cases, in order to highlight the impact of margining and funding costs in pricing a deal.

\section[Bilateral Collateralized Credit and Funding Valuation Adjusted Price]{Bilateral Collateralized Credit and Funding\\Valuation Adjusted Price}
\label{sec:BCCFVA}

Here, we develop a risk-neutral evaluation methodology for Bilateral Collateralized Credit Valuation Adjusted (BCCVA) price which we extend to cover the case of Bilateral Collateralized Credit and Funding Valuation Adjusted (BCCFVA) price by including funding costs. Along the way, we highlight the relevant market standards and agreements which we follow to derive such formulas. We refer the reader to \cite{BrigoCapponiPallaviciniPapatheodorou} for an extensive discussion of market considerations and of collateral mechanics, which also includes an analysis of credit valuation adjustments on interest rate swaps in presence of different collateralization strategies.

We assume that to price a derivative we have to consider all the cash flows occurring when the trading position is entered. We can group them as follows
\begin{enumerate}
\item derivative cash flows (e.g. coupons, dividends, etc\dots) inclusive of hedging instruments;
\item cash flows required by the collateral margining procedure;
\item cash flows required by the funding and investing procedures;
\item cash flows occurring on default events.
\end{enumerate}%
Notice that we discount cash flows by using the risk-free discount factor $D(t,T)$, since all costs are included as additional cash flows rather than ad hoc spreads.

We refer to the two names involved in the financial contract and subject to default risk as \textit{investor} (also called name ``0'') and \textit{counterparty } (also called name ``2''). In cases where the portfolio exchanged by the two parties is also a default sensitive instrument, we introduce a third name referring to the underlying reference credit of that portfolio (also called name ``1'').

We denote by $\tau_I$,and $\tau_C$ respectively the default times of the investor and counterparty. We fix the portfolio time horizon $T \in \mathds{R}^+$, and fix the risk neutral pricing model $(\Omega,\mathcal{G},\mathbb{Q})$, with a filtration $(\mathcal{G}_t)_{t \in [0,T]}$ such that $\tau_C$, $\tau_I$ are $\mathcal{G}$-stopping times. We denote by $\Exx_t$ the conditional expectation under $\Qxx$ given $\mathcal{G}_t$, and by $\Exx_{\tau_i}$ the conditional expectation under $\Qxx$ given the stopped filtration $\mathcal{G}_{\tau_i}$. We exclude the possibility of simultaneous defaults, and define the first default event between the two parties as the stopping time
\[
\tau := \tau_C \wedge \tau_I.
\]

The main result of the present paper is the pricing equation (BCCFVA price) for a deal inclusive of counterparty credit risk (CVA and DVA), margining costs, and funding and investing costs.

The BCCFVA price ${\bar V}_t$ of a derivative contract, which is derived in the following sections, is given by
\begin{eqnarray}
{\bar V}_t(C;F) & = & \Ex{t}{\Pi(t,T\wedge\tau) + \gamma(t,T\wedge\tau;C) + \varphi(t,T\wedge\tau;F) } \\\nonumber
                & + & \Ex{t}{\ind{\tau<T} D(t,\tau) \theta_\tau(C,\varepsilon) },
\end{eqnarray}%
where
\begin{itemize}
\item $\Pi(t,T)$ is the sum of all discounted payoff terms in the interval $(t,T]$,
\item $\gamma(t,T;C)$ are the collateral margining costs within such interval, $C$ being the collateral account,
\item $\varphi(t,T;F)$ the funding and investing costs within such interval, $F$ being the cash account needed for trading, and
\item $\theta_\tau(C,\varepsilon)$ the on-default cash flow, $\varepsilon$ being the amount of losses or costs the surviving party would incur upon a default event (close-out amount).
\end{itemize}

The margining procedure and the liquidity policy dictate respectively the dynamics of the collateral account $C_t$ and of the cash account $F_t$, while the close-out amount $\varepsilon_t$ is defined by the CSA holding between the counterparties. Common strategies, as we will see later on, may link the values of such processes to the price of the derivative itself, transforming the previous definition into a {\em recursive} equation. This feature is hidden in simplified approaches based on adding a spread to the discount curve to accommodate collateral and funding costs. A different approach is followed by \cite{Crepey2011} and \cite{BurgardKjaer2011} where the usual risk-neutral evaluation framework is extended to include many cash accounts accruing at different rates. Yet, a similar structure for the derivative price is obtained as a solution of a backward SDE.

In the following sections we expand all the above terms to allow the calculation of the BCCFVA price.

\section{Trading under ISDA Master Agreement}
\label{sec:ISDA}

The ISDA Master Agreement lists two different tools to reduce counterparty credit risk: collateralization by a margining procedure and close-out netting rules. Both tools are ruled by the Credit Support Annex (CSA) holding between the counterparties of the deal. 

Collateralization means the right of recourse to some asset of value that can be sold or the value of which can be applied as a guarantee in the event of default on the transaction. Close-out netting rules apply when a default occurs, and force multiple obligations towards a counterparty to be consolidated into a single net obligation before recovery is applied.

Here, we briefly describe these tools to analyze the margining costs required by the collateral posting, and to integrate their price within our subsequent funding costs model, leading eventually to calculate the Bilateral Collateralized Credit Valuation Adjusted (BCCVA) price ${\bar V}_t(C;0)$.

\subsection{Re-Hypothecation Liquidity Risk}

In case of no-default happening, at final maturity the collateral provider expects to get back the remaining collateral from the collateral taker. Similarly, in case of default happening earlier (and assuming the collateral taker before default to be the surviving party), after netting the collateral account with the cash flows of the transaction, the collateral provider expects to get back the remaining collateral on the account if any. However, it is often considered to be important, commercially, for the collateral taker to have relatively unrestricted use of the collateral until it must be returned to the collateral provider. Thus, when re-hypothecation occurs, the collateral provider must consider the possibility to recover only a fraction of his collateral, see \cite{BrigoCapponiPallaviciniPapatheodorou} for the details. 

If the investor is the collateral taker, we denote the recovery fraction on the collateral re-hypothecated by the defaulted investor by $\rec'_0$, while if the counterparty is the collateral taker, then we denote the recovery fraction on collateral re-hypothecated by the counterparty by $\rec'_2$. Accordingly, we define the collateral loss incurred by the counterparty upon investor default by $\lgd'_0 = 1 - \rec'_0$ and the collateral loss incurred by the investor upon counterparty default by $\lgd'_2 = 1 - \rec'_2$.

Typically, the surviving party has precedence on other creditors to get back his collateral, so that $\rec_0 \leq \rec_0' \leq 1$, and $\rec_2 \leq \rec_2' \leq 1$. Here, $\rec_0$ ($\rec_2$) denote the recovery fraction of the market value of the transaction that the counterparty (investor) gets when the investor (counterparty) defaults. Notice that the case when collateral cannot be re-hypothecated and has to be kept into a segregated account is obtained by setting $\rec_0' = \rec_2' = 1$.

The impact of re-hypothecation on the pricing of counterparty risk is analyzed in \cite{BrigoCapponiPallaviciniPapatheodorou}.

\subsection{Collateral Management under Margining Procedure}

A margining procedure consists in a pre-fixed set of dates during the life of a deal when both parties post or withdraw collaterals, according to their current exposure, to or from an account held by the Collateral Taker. A realistic margining practice should allow for collateral posting only on a fixed time-grid ($t_1,\ldots,t_n$), and for the presence of independent amounts, minimum transfer amounts, thresholds, and so on, as described in \cite{BrigoCapponiPallaviciniPapatheodorou}. Here, we present a framework which does not rely on the particular margining procedure adopted by the counterparties.

Without loss of generality, we assume that the collateral account $C_t$ is held by the investor if $C_t>0$, and by the counterparty if $C_t<0$. If at time $t$ the investor posts some collateral we write that $dC_t<0$, and the other way round if the counterparty is posting.

The CSA agreement holding between the counterparties ensures that the collateral taker remunerates the account at a particular accrual rate. We introduce the (forward) collateral accrual rates, namely $c^+_t(T)$ when collateral assets are taken by the investor, and $c^-_t(T)$ in the other case, as adapted processes. Further, we define the (collateral) zero-coupon bonds $P^{c^\pm}_t(T)$ as given by
\[
P^{c^\pm}_t(T) := \frac{1}{1+(T-t)c^\pm_t(T)}.
\]

We assume that interests accrued by the collateral account are saved into the account itself, so that they can be directly included into close-out and margining procedures. Thus, any cash-flow due to collateral costs or accruing interests can be dropped from our list, since it can be considered as a flow within each counterparty.

We start by listing all cash-flows originating from the investor and going to the counterparty if default events do not occur:
\begin{enumerate}
\item the investor opens the account at the first margining date $t_1$ if $C_{t_1}<0$;
\item he posts to or withdraws from the account at each $t_k$ as long as $C_{t_k}<0$ by considering a collateral account's growth at CSA rate $c^-_{t_k}(t_{k+1})$;
\item he closes the account at the last margining date $t_m$ if $C_{t_m}<0$.
\end{enumerate}
The counterparty considers the same cash-flows for opposite values of the collateral account at each margining date. Hence, we can sum all such contributions. If we do not take into account default events, we define the sum of all (discounted) margining cash flows as given by
\begin{eqnarray*}
\Gamma(t,T;C)
 & := & \,\ind{t\le t_1<T} C^-_{t_1} D(t,t_1) - \ind{t<t_n\le T} C^-_{t_n} D(t,t_n) \\
 &  - & \sum_{k=1}^{n-1} \ind{t<t_{k+1}\le T} \left( C^-_{t_k} \frac{1}{P^{c^-}_{t_k}(t_{k+1})} - C^-_{t_{k+1}} \right) D(t,t_{k+1}) \\
 &  + & \,\ind{t\le t_1<T} C^+_{t_1} D(t,t_1) - \ind{t<t_n\le T} C^+_{t_n} D(t,t_n) \\
 &  - & \sum_{k=1}^{n-1} \ind{t<t_{k+1}\le T} \left( C^+_{t_k} \frac{1}{P^{c^+}_{t_k}(t_{k+1})} - C^+_{t_{k+1}} \right) D(t,t_{k+1}),
\end{eqnarray*}%
where in our notation $X^+ := \max(X,0)$ and $X^- := \min(X,0)$. Then, to introduce default events, we can simply stop collateral margining when they occur, so that we have
\[
{\bar\Gamma}(t,T;C) := \Gamma(t,T\wedge\tau;C),
\]%
where $\bar\Gamma$ is the sum of all (discounted) margining cash flows up to the first default event. 

We can re-arrange the previous definitions, by taking expectation the under risk-neutral measure, and we obtain
\[
\Ex{t}{{\bar\Gamma}(t,T;C)} = \Ex{t}{\gamma(t,T\wedge\tau;C) + \ind{\tau<T} D(t,\tau) C_{\tau^-} },
\]%
where the margining costs are defined as
\begin{equation}
\gamma(t,T;C) :=
 \sum_{k=1}^{n-1} \ind{t\le t_k<T} D(t,t_k) \left( C_{t_k} - C^-_{t_k} \frac{P_{t_k}(t_{k+1})}{P^{c^-}_{t_k}(t_{k+1})} - C^+_{t_k} \frac{P_{t_k}(t_{k+1})}{P^{c^+}_{t_k}(t_{k+1})} \right),
\end{equation}%
and we introduce the pre-default value $C_{\tau^-}$ of the collateral account as given by
\begin{equation}
C_{\tau^-} := \sum_{k=1}^{n-1} \ind{t_k<\tau<t_{k+1}} \left( C^-_{t_k} \frac{P_\tau(t_{k+1})}{P^{c^-}_{t_k}(t_{k+1})} + C^+_{t_k} \frac{P_\tau(t_{k+1})}{P^{c^+}_{t_k}(t_{k+1})} \right).
\end{equation}%

Notice that the pre-default value $C_{\tau^-}$ of the collateral account is used by the CSA to calculate close-out netted exposures, it can be different from the actual value of the collateral account at the default event, since some collateral assets (or all) might be re-hypothecated.

\subsection{Close-Out Netting Rules}

The occurrence of a default event gives the parties the right to terminate all transactions that are included under the relevant ISDA Master Agreement. The ISDA Master Agreement provides for the mechanism of close-out netting to be enforced. The surviving party should evaluate the terminated transactions to claim for a reimbursement after the application of netting rules consolidating the transactions, inclusive of collateral accounts.

The ISDA Master Agreement defines the term {\em close-out amount} to be the amount of losses or costs the surviving party would incur in replacing or in providing for an economic equivalent to the relevant transaction. Notice that the close-out amount is not a symmetric quantity w.r.t.\ the exchange of the role of the two parties, since it is valued by one party after the default of the other one.

We introduced the close-out amount in section \ref{sec:BCCFVA} as one of the elements needed to calculate on-default cash flows, and we named it $\varepsilon$. Since its value depends on which is the surviving party, we specialize it in the following form:
\[
\varepsilon_\tau := \ind{\tau=\tau_C} \varepsilon_{I,\tau} + \ind{\tau=\tau_I} \varepsilon_{C,\tau},
\]
where we define the close-out amount priced at time $\tau$ by the investor on counterparty's default with $\varepsilon_{I,\tau}$ (and $\varepsilon_{C,\tau}$ in the other case, namely when the investor is defaulting). Notice that we always consider all prices from the point of view of the investor. Thus, a positive value for $\varepsilon_{I,\tau}$ means the investor is a creditor of the counterparty, while a negative value for $\varepsilon_{C,\tau}$ means the counterparty is a creditor of the investor. The ISDA documentation is not very tight in defining how one has to calculate the close-out amounts, and it can clearly produce a wide range of results as described in \cite{Parker2009} and \cite{Weeber2009}. We refer to \cite{BrigoCapponiPallaviciniPapatheodorou} and to references therein for further discussion.

By following \cite{BrigoCapponiPallaviciniPapatheodorou}, with a view at extending it, we start by listing all the situations that may arise on counterparty default and investor default events. Our goal is to calculate the present value of all cash flows involved by the contract by taking into account (i) collateral margining operations, and (ii) close-out netting rules in case of default. Notice that we can safely aggregate the cash flows of the contract with the ones of the collateral account, since on contract termination all the posted collateral is returned to the originating party.

We start by considering all possible situations which may arise at the default time of the counterparty, which is assumed to default before the investor. We have:

\begin{enumerate}

\item The investor measures a positive (on-default) exposure on counterparty default ($\varepsilon_{I,\tau_C}>0$), and some collateral posted by the counterparty is available ($C_{\tau_C-}>0$). Then, the investor exposure is reduced by netting, and the remaining collateral (if any) is returned to the counterparty. If the collateral is not enough, the investor suffers a loss for the remaining exposure. Thus, we have
\[
\ind{\tau=\tau_C<T} \ind{\varepsilon_{\tau} > 0} \ind{C_{\tau-}>0} (\rec_C(\varepsilon_{\tau} - C_{\tau-})^+ + (\varepsilon_{\tau}-C_{\tau-})^-).
\]

\item The investor measures a positive (on-default) exposure on counterparty default ($\varepsilon_{I,\tau_C}>0$), and some collateral posted by the investor is available ($C_{\tau_C-}<0$). Then, the investor suffers a loss for the whole exposure. All the collateral (if any) is returned to the investor if it is not re-hypothecated, otherwise only a recovery fraction of it is returned. Thus, we have
\[
\ind{\tau=\tau_C<T} \ind{\varepsilon_{\tau} > 0} \ind{C_{\tau-}<0} (\rec_C \varepsilon_{\tau} - \rec'_C C_{\tau-}).
\]

\item The investor measures a negative (on-default) exposure on counterparty default ($\varepsilon_{I,\tau_C}<0$), and some collateral posted by the counterparty is available ($C_{\tau_C-}>0$). Then, the exposure is paid to the counterparty, and the counterparty gets back its collateral in full, yielding
\[
\ind{\tau=\tau_C<T} \ind{\varepsilon_{\tau} < 0} \ind{C_{\tau-}>0} (\varepsilon_{\tau} - C_{\tau-}).
\]

\item The investor measures a negative (on-default) exposure on counterparty default ($\varepsilon_{I,\tau_C}<0$), and some collateral posted by the investor is available ($C_{\tau_C-}<0$). Then, the exposure is reduced by netting and paid to the counterparty. The investor gets back its remaining collateral (if any) in full if it is not re-hypothecated, otherwise he only gets the recovery fraction of the part of collateral exceeding the exposure, leading to
\[
\ind{\tau=\tau_C<T} \ind{\varepsilon_{\tau} < 0} \ind{C_{\tau-}<0} ( (\varepsilon_{\tau} - C_{\tau-})^- + \rec'_C (\varepsilon_{\tau} - C_{\tau-})^+ ).
\]

\end{enumerate}

Similarly, if we consider all possible situations which can arise at the default time of the investor, and then aggregate all these cash flows, along with the ones due in case of non-default, inclusive of the collateral account, we obtain, after straightforward manipulations
\[
\begin{split}
& {\bar \Pi}(t,T;C) := \\
& \quad \quad \,\,\! \Pi(t,T\wedge\tau) + {\bar\Gamma}(t,T;C) \\
& \quad + \ind{\tau=\tau_C<T} D(t,\tau) \ind{\varepsilon_{I,\tau}<0} \ind{C_{\tau^-}>0} (\varepsilon_{I,\tau} - C_{\tau^-}) \\
& \quad + \ind{\tau=\tau_C<T} D(t,\tau) \ind{\varepsilon_{I,\tau}<0} \ind{C_{\tau^-}<0} ((\varepsilon_{I,\tau} - C_{\tau^-})^- + \rec'_C (\varepsilon_{I,\tau} - C_{\tau^-})^+) \\
& \quad + \ind{\tau=\tau_C<T} D(t,\tau) \ind{\varepsilon_{I,\tau}>0} \ind{C_{\tau^-}>0} ((\varepsilon_{I,\tau} - C_{\tau^-})^- + \rec_C(\varepsilon_{I,\tau} - C_{\tau^-})^+) \\
& \quad + \ind{\tau=\tau_C<T} D(t,\tau) \ind{\varepsilon_{I,\tau}>0} \ind{C_{\tau^-}<0} (\rec_C \varepsilon_{I,\tau} - \rec'_C C_{\tau^-}) \\
& \quad + \ind{\tau=\tau_I<T} D(t,\tau) \ind{\varepsilon_{C,\tau}>0} \ind{C_{\tau^-}<0} (\varepsilon_{C,\tau} - C_{\tau^-}) \\
& \quad + \ind{\tau=\tau_I<T} D(t,\tau) \ind{\varepsilon_{C,\tau}>0} \ind{C_{\tau^-}>0} ((\varepsilon_{C,\tau} - C_{\tau^-})^+ + \rec'_I (\varepsilon_{C,\tau} - C_{\tau^-})^-) \\
& \quad + \ind{\tau=\tau_I<T} D(t,\tau) \ind{\varepsilon_{C,\tau}<0} \ind{C_{\tau^-}<0} ((\varepsilon_{C,\tau} - C_{\tau^-})^+ + \rec_I(\varepsilon_{C,\tau} - C_{\tau^-})^-) \\
& \quad + \ind{\tau=\tau_I<T} D(t,\tau) \ind{\varepsilon_{C,\tau}<0} \ind{C_{\tau^-}>0} (\rec_I \varepsilon_{C,\tau} - \rec'_I C_{\tau^-}).
\end{split}
\]%

We define the Bilateral Collateralized Credit Valuation Adjusted (BCCVA) price ${\bar V}_t(C;0)$, without considering funding and investing costs, by taking the risk-neutral expectation of the previous equation, and we get
\begin{eqnarray}
\label{eq:bccva}
{\bar V}_t(C;0)
  & := & \Ex{t}{{\bar\Pi}(t,T;C)} \\\nonumber
  &  = & \Ex{t}{\Pi(t,T\wedge\tau) + \gamma(t,T\wedge\tau;C) + \ind{\tau<T} D(t,\tau) \theta_\tau(C,\varepsilon) },
\end{eqnarray}%
where we define the on-default cash flow $\theta_{\tau}(C,\varepsilon)$ as given by
\begin{eqnarray}
\label{eq:theta}
\theta_{\tau}(C,\varepsilon)
 & := & \,\ind{\tau=\tau_C<\tau_I} \left( \varepsilon_{I,\tau} - \lgd_C (\varepsilon_{I,\tau}^+ - C_{\tau^-}^+)^+ - \lgd'_C (\varepsilon_{I,\tau}^- - C_{\tau^-}^-)^+ \right) \\\nonumber
 &  + & \,\ind{\tau=\tau_I<\tau_C} \left( \varepsilon_{C,\tau} - \lgd_I (\varepsilon_{C,\tau}^- - C_{\tau^-}^-)^- - \lgd'_I (\varepsilon_{C,\tau}^+ - C_{\tau^-}^+)^- \right),
\end{eqnarray}%
namely to price a deal (without funding costs) we have to sum up three components: (i) deal's cash flows, (ii) margining costs, and (iii) close-out amount reduced by the CVA/DVA contribution.

\subsubsection{Perfect Collateralization}

As an example of the BCCVA pricing formula we consider the case of perfect collateralization, which we define as given by collateralization in continuous time, with continuous mark-to-market of the portfolio at default events, and with collateral account inclusive of margining costs at any time $u$, \emph{i.e.}
\[
C_u \doteq \Ex{u}{\Pi(u,T) + \gamma(u,T;C)},
\]%
with close-out amount equal to collateral price, \emph{i.e.}
\[
\varepsilon_{I,\tau} \doteq \varepsilon_{C,\tau} \doteq C_\tau.
\]%

Then, from the BCCVA price equation we get
\begin{eqnarray*}
{\bar V}_t(C;0) & = & \Ex{t}{\Pi(t,T\wedge\tau) + \gamma(t,T\wedge\tau;C) + \ind{\tau<T} D(t,\tau) C_\tau } \\
                & = & \Ex{t}{\Pi(t,T) + \gamma(t,T;C) } \\
                & = & C_t.
\end{eqnarray*}%
Now, we consider two margining dates $t_k$ and $t_{k+1}$. By substituting the expression for margining cash flows we get (up to maturity)
\[
{\bar V}^\pm_{t_k}(C;0) = \frac{P^{c^\pm}_{t_k}(t_{k+1})}{P_{t_k}(t_{k+1})} \left( \Ex{t_k}{D(t_k,t_{k+1}){\bar V}_{t_{k+1}}(C;0) +\Pi(t_k,t_{k+1})} \right)^\pm.
\]%

We can continue the example for the case of symmetric CSA rates, where we get, with $t_1=t$,
\[
{\bar V}_t(C;0) = \Ex{t}{ \sum_{k=1}^{n-1} \left( \prod_{i=1}^k \frac{P^c_{t_i}(t_{i+1})}{P_{t_i}(t_{i+1})} \right) \Pi(t_k,t_{k+1}) D(t,t_k) }.
\]%
Then, taking the limit of continuous collateralization, we get
\[
{\bar V}_t(C;0) = \Ex{t}{ \int_t^T \Pi(u,u+du) \, \exp\left\{ -\int_t^u dv\, c_v \right\} },
\]%
where the (symmetric) instantaneous CSA rate is defined as $c_t:=c^\pm_t(t)$. Thus, we have that perfect collateralization with symmetric CSA rates means {\em collateral discounting}.

\section{Funding Risk and Liquidity Policies}
\label{sec:funding}

We start our discussion by referring to a working paper of the Basel Committee, {\it``International Framework for Liquidity Risk Measurement, Standards and Monitoring''} of December 2009, that investigates market and funding liquidity issues. We will not proceed with its level of generality here, since we will be dealing mostly with pricing and funding as related to CVA. In pricing applications, modeling consistently funding costs with bilateral CVA and collateral margining is a complex task, since it includes modelling the bank's liquidity policy, and to some extent the banking system as a whole.

In this respect, realistic funding liquidity modelling can be found in the literature, see for instance \cite{Drehmann2010}, or \cite{Brunnermeier2009} and references therein. Yet, here we resort to risk-neutral evaluation of funding costs, in a framework similar for example to the one by \cite{Crepey2011}. More examples can be found in \cite{BurgardKjaer2011}, \cite{MoriniPrampolini2011}, \cite{Fries2010}, or \cite{Pallavicini2010}.

In practice, this means that while not addressing the details of funding liquidity modelling, we can simply introduce risk-neutral funding costs in terms of additional costs needed to complete each cash-flow transaction. We may add cost of funding along with (bilateral) credit valuation adjustments by collecting all costs coming from funding the trading position, inclusive of hedging costs, and collateral margining. We may have some asymmetries here, since the prices calculated by one party may differ from the same ones evaluated by the other party, since each price contains only funding costs undertaken by the calculating party.

\subsection{Funding, Hedging and Collateralization}

Without entering into detail on funding liquidity modelling, we can introduce risk-neutral funding costs by considering the positions entered by the trader at time $t$ to obtain the amount of cash $F_t>0$ needed to establish the hedging strategy, along with the positions used to invest cash surplus $F_t<0$. If the deal is collateralized we include the margining procedure into the deal definition, so that we are able to evaluate also its funding costs. Notice that, if collateral re-hypothecation is allowed, each party may use the collateral account $C_t$ for funding, and, if the collateral account is large enough, we could drop all funding costs. See \cite{Fujii2010}.

In order to write an explicit formula for cash flows we need an expression for the cash amount $F_t$ to be funded or invested. Such problem is faced also in \cite{Piterbarg2010}, \cite{BurgardKjaer2011} and \cite{Crepey2011}. 

We notice that the hedging strategy, perfectly replicating the derivative to be priced, is formed by a cash amount, namely our cash account $F_t$, and a portfolio $H_t$ of hedging instruments, so that, if the deal is not collateralized, or re-hypothecation is forbidden, we get
\[
F_t = {\bar V}_t(0;F) - H_t,
\]%
where ${\bar V}_t(0;F)$ is the derivative risky price inclusive of funding and investing costs.

On the other hand, if re-hypothecation is allowed we can use collateral assets for funding, so that the amount of cash needed to be funded or invested is reduced and it is given by
\[
F_t = {\bar V}_t(C;F) - C_t - H_t,
\]%
where ${\bar V}_t(C;F)$ is the derivative risky price inclusive of funding and investing costs.

We do not add hedging costs explicitly (e.g. lending or convenience rates), since we can deal with them by modifying the drifts of the underlying assets. 

Notice that we obtain a {\em recursive} equation, since the derivative price at time $t$ depends on the funding strategy after $t$, and in turn the funding strategy after $t$ will depend on the derivative price at following times. This will be made explicit in the following sections.

\subsection{Liquidity Policies}

The positions entered by the trader for funding or investing depend on his liquidity policy, namely we assume that any cash amount $F_t>0$ needed by the trader, or any cash surplus $F_t<0$ to be invested, can be managed by entering a position with an external party, for instance the treasury or a funder operating on the market.

In particular, we consider that the trader enters a funding position according to a time-grid $t_1,\ldots,t_m$. More precisely, between two following grid times $t_j$ and $t_{j+1}$ we have that
\begin{itemize}
\item at $t_j$ the trader asks the funder for a cash amount equal to $F_{t_j}$,
\item at $t_{j+1}$ the trader has to reimburse the funder for the cash amount previously obtained and has to pay for funding costs.
\end{itemize}%
Further, we assume that funding costs are established at the starting date of each funding period and charged at the end of the same period. We can follow the same line of reasoning also for investing cash amounts ($F_t<0$) not directly used by the trader, and to consider investing periods along with funding periods.

The price of funding and investing contracts may be introduced without loss of generality as an adapted process $P^{f^+}_t(T)$, measurable at $t$, representing the price of a funding contract where the trader pays one unit of cash at maturity date $T>t$, and the price $P^{f^-}_t(T)$ of an investing contract where the trader receives one unit of cash at maturity date. We introduce also the (forward) funding and investing rates
\[
f^\pm_t(T) := \frac{1}{T-t}\left(\frac{1}{P^{f^\pm}_t(T)}-1\right).
\]

Hence, we can define the derivative price ${\bar V}_t(C;F)$ inclusive of funding costs and collateral management as given by
\[
{\bar V}_t(C;F) := \Ex{t}{ {\bar\Pi(t,T;C)} + \varphi(t,T\wedge\tau;F) },
\]%
where $\varphi(t,T\wedge\tau;F)$ is the sum of discounted cash flows coming from all the funding and investing positions opened by the investor to hedge its trading position according to his liquidity policy up to first default event, while ${\bar\Pi}(t,T;C)$ is the sum of discounted cash flows coming from contract payout, collateral margining procedure, and close-out netting rules, as given in the previous section.

Before defining $\varphi(t,T\wedge\tau;F)$, we describe two examples of of liquidity policies to better understand our approach to funding and investing costs. Indeed, we distinguish between funding and investing in terms of returns since in general $P^{f^-}$ and $P^{f^+}$ will be different. We may call this approach the ``micro'' approach to funding, or possibly the bottom-up approach. This is not the unique possibility: one could assume an average cost of funding to be applied to all deals, and an average return for investing. This would lead to two curves for $P^{f^-}$ and $P^{f^+}$ that would hold for all funding costs and invested amounts respectively, regardless of the specific deal.

On the other hand, one can go further and assume that the cost of investing and the cost of funding match, so that $P^{f^-}$ and $P^{f^+}$ are not only the same across deals, but are equal to each other, implying a common funding and investing spread. In practice the spread would be set at a common value for borrowing or lending, and this value would match what goes on across all deals on average. This would be a ``large-pool'' or homogeneous average approach, which would look at a unique funding spread for the bank to be applied to all products traded by capital markets. 

\subsubsection{Funding via Bank's Treasury}

\begin{figure}
\begin{center}
\scalebox{0.9}{
\begin{tikzpicture}
   \node [trader, node distance=2cm] (I1) {Trader $1$};
   \node [below of=I1, node distance=2cm] (I2) {$\vdots$};
   \node [trader, below of=I2, node distance=2cm] (IN) {Trader $N$};
   \node [bank, left of=I2, text height=6cm,node distance = 5cm] (T) {Treasury};
   \node [market, left of=T, text height=6cm, node distance = 8cm] (M) {Market};
   \path [line] (T.east)+(0,2) -- node [near end,above] {$f^{+,I_1}_t$} (I1);
   \path        (T.east)+(0,2) -- node [near start,below] {$f^{-,I_1}_t$} (I1);
   \path [line] (T.east)+(0,-2) -- node [near end,above] {$f^{+,I_N}_t$} (IN);
   \path        (T.east)+(0,-2) -- node [near start,below] {$f^{-,I_N}_t$} (IN);
   \path [line] (M.east) -- node [near start,below] {$r_t+\lambda^F_t+\ell^-_t$} (T);
   \path        (M.east) -- node [near end,above] {$r_t+\lambda^I_t+\ell^+_t$} (T);
   \begin{pgfonlayer}{background}
   \node [background, fit=(I1) (IN) (T), label=above:Bank] {};
   \end{pgfonlayer}
\end{tikzpicture}}
\end{center}
\caption{Traders may fund and invest only by means of their treasury. Thus, average rates $f^\pm$ are applied. The funding and investing trades closed by treasury are not seen by the traders.}
\label{fig:ftp}
\end{figure}
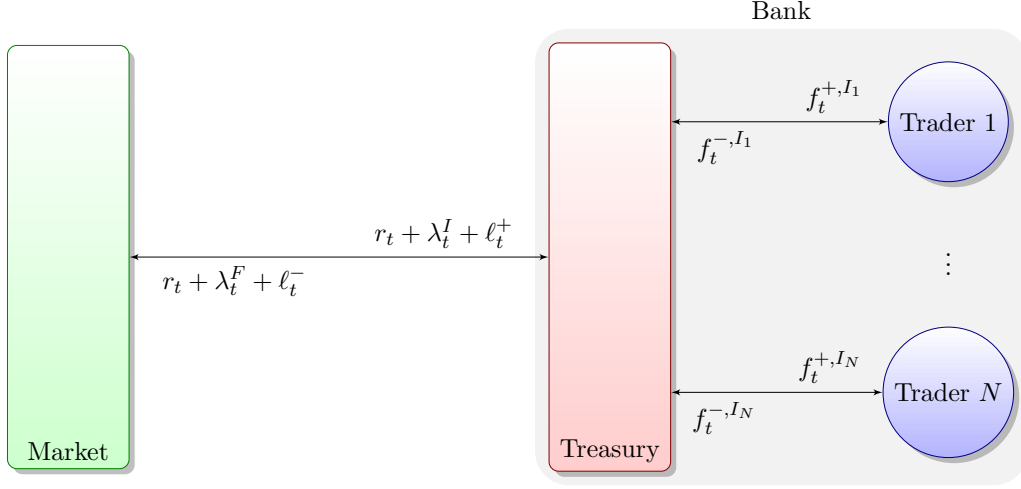

As a first example, we can consider that the counterparty of funding and investing cash flows is the treasury, which in turn operates on the market. Thus, funding and investing rates $f^\pm_t$ for each trader are determined by the treasury, for instance by means of a funds transfer pricing (FTP) process which allows to measure the performance of different business units. A pictorial representation is given in figure \ref{fig:ftp}.

From the point of view of the investor the following (discounted) cash flows occur when entering a funding or investing position $\Phi_j$ at $t_j$:
\[
\Phi_j(t_j,t_{j+1};F) := F_{t_j} - N_{t_j} D(t_j,t_{j+1}),
\]%
with
\[
N_{t_j} := \frac{F^-_{t_j}}{P^{f^-}_{t_j}(t_{j+1})} + \frac{F^+_{t_j}}{P^{f^+}_{t_j}(t_{j+1})}.
\]%
The investor does not operate directly on the market, but only with his treasury. Thus, in case of default both the parties of the funding/investing deal disappear, without any further cash flow. In particular, in such case the treasury, and not the trader, is in charge of debt valuation adjustments due to funding positions, so that we can consider that the funding/investing is in place only if default events do not happen, leading to the following definition of the funding/investing cash flows when default events are considered:
\[
{\bar \Phi}_j(t_j,t_{j+1};F) := \ind{\tau>t_j} \Phi_j(t_j,t_{j+1};F).
\]%

Thus, the price of all cash flows coming from funding and investing positions is given by
\begin{eqnarray*}
\varphi(t,T;F) &:=& \sum_{j=1}^{m-1} {\bar \Phi}_j(t_j,t_{j+1};F) \\
               & =& \sum_{j=1}^{m-1} \ind{\tau>t_j} \left( F_{t_j} - F^-_{t_j} \frac{P_{t_j}(t_{j+1})}{P^{f^-}_{t_j}(t_{j+1})} - F^+_{t_j} \frac{P_{t_j}(t_{j+1})}{P^{f^+}_{t_j}(t_{j+1})} \right).
\end{eqnarray*}

\subsubsection{Funding Directly on the Market}

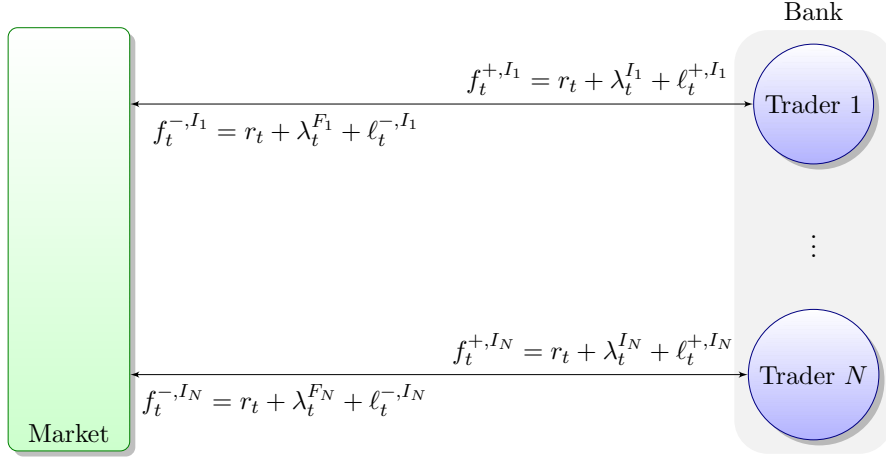
\begin{figure}
\begin{center}
\scalebox{0.9}{
\begin{tikzpicture}
   \node [trader, node distance=2cm] (I1) {Trader $1$};
   \node [below of=I1, node distance=2cm] (I2) {$\vdots$};
   \node [trader, below of=I2, node distance=2cm] (IN) {Trader $N$};
   \node [market, left of=I2, text height=6cm, node distance = 11cm] (M) {Market};
   \path [line] (M.east)+(0,2) -- node [near end,above] {$f^{+,I_1}_t=r_t+\lambda^{I_1}_t+\ell^{+,I_1}_t$} (I1);
   \path        (M.east)+(0,2) -- node [near start,below] {$f^{-,I_1}_t=r_t+\lambda^{F_1}_t+\ell^{-,I_1}_t$} (I1);
   \path [line] (M.east)+(0,-2) -- node [near end,above] {$f^{+,I_N}_t=r_t+\lambda^{I_N}_t+\ell^{+,I_N}_t$} (IN);
   \path        (M.east)+(0,-2) -- node [near start,below] {$f^{-,I_N}_t=r_t+\lambda^{F_N}_t+\ell^{-,I_N}_t$} (IN);
   \begin{pgfonlayer}{background}
   \node [background, fit=(I1) (IN), label=above:Bank] {};
   \end{pgfonlayer}
\end{tikzpicture}}
\end{center}
\caption{Traders may fund and invest directly on the market. Thus, funding and investing rates $f^\pm$ must match the market rates. Here $\lambda$ are the default intensities of traders or funder, and $\ell^\pm$ the liquidity (bond/CDS) basis for buying or selling.}
\label{fig:dm}
\end{figure}

As a second example of liquidity policy, we can consider that each trader operates directly on the market to enter funding and investing positions (see \cite{Crepey2011}). Here, the treasury has no longer an active role, and it could be dropped from our scheme, as shown in figure \ref{fig:dm}

The investor operates directly on the market. Thus, his mark-to-market should include default debt valuation adjustments due to funding positions. Here, we consider the funder to be risk-free.

By using the BCCVA pricing formula without collateralization, we obtain the sum of the funding (discounted) cash flows inclusive of debt valuation adjustments, as given by
\begin{eqnarray*}
{\bar\Phi}_j(t_j,t_{j+1};F)
 & := & \ind{\tau>t_j} \Phi_j(t_j,t_{j+1}\wedge\tau_I;F) \\
 &  - & \ind{\tau_I<t_{j+1}} (\lgd_I \varepsilon^-_{F,\tau_I} - \varepsilon_{F,\tau_I}) D(t_j,\tau_I),
\end{eqnarray*}%
where $\varepsilon_{F,t}$ is the close-out amount calculated by the funder on investor's default event, which we assume to be
\[
\varepsilon_{F,\tau_I} := - N_{t_j} P_{\tau_I}(t_{j+1}).
\]

Notice that, by following \cite{Crepey2011}, we could assume a recovery rate for the investor different from the one we use as recovery rate for trading deals, since the seniority could be different. It is straightforward to extend the present case in such direction.

Thus, the price of all cash flows coming from funding and investing positions is given by
\begin{eqnarray*}
\varphi(t,T;F) &:=& \sum_{j=1}^{m-1} {\bar \Phi}_j(t_j,t_{j+1};F) \\
               & =& \sum_{j=1}^{m-1} \ind{\tau>t_j} \left( F_{t_j} - F^-_{t_j} \frac{P_{t_j}(t_{j+1})}{P^{f^-}_{t_j}(t_{j+1})} - F^+_{t_j} \frac{P_{t_j}(t_{j+1})}{{\bar P}^{f^+}_{t_j}(t_{j+1})} \right),
\end{eqnarray*}%
where the risky-adjusted funding zero-coupon bond ${\bar P}^{f^+}_t(T)$ is defined as
\[
{\bar P}^{f^+}_t(T) := \frac{P^{f^+}_t(T)}{ \Ex{t}{ \lgd_I\ind{\tau_I>T} + \rec_I } }.
\]

Notice that, if we set $\rec_I=1$ (so that $\lgd_I=0$) we get ${\bar P}^{f^+}_t(T)$ being equal to $P^{f^+}_t(T)$, and we recover the previous example. Thus, in the following we simply write $P^{f^\pm}_t(T)$ to avoid cumbersome notations.

\subsection{BCCVA Pricing with Funding Costs (BCCFVA)}

In this paper we stay as general as possible and therefore assume the micro view, but of course it is enough to collapse our variables to common values to obtain any of the large-pool approaches. By the previous examples we understand that a sensible choice for funding and investing cash flows could be
\[
\varphi(t,T;F) :=
 \sum_{j=1}^{m-1} \ind{t\le t_j<T} D(t,t_j) \left( F_{t_j} - F^-_{t_j} \frac{P_{t_j}(t_{j+1})}{P^{f^-}_{t_j}(t_{j+1})} - F^+_{t_j} \frac{P_{t_j}(t_{j+1})}{P^{f^+}_{t_j}(t_{j+1})} \right),
\]
whatever the definition of funding and investing rates may be.

Hence, the BCCFVA price ${\bar V}_t(C;F)$, inclusive of funding and investing costs, can be written in the following form:
\begin{eqnarray}
\label{eq:bccfva}
{\bar V}_t(C;F) & = & \Ex{t}{\Pi(t,T\wedge\tau) + \gamma(t,T\wedge\tau;C) + \varphi(t,T\wedge\tau;F) } \\\nonumber
                & + & \Ex{t}{\ind{\tau<T} D(t,\tau) \theta_\tau(C,\varepsilon) }.
\end{eqnarray}%

The above formula, when funding and margining costs are discarded, collapses to the formula of BCCVA adjusted price found in \cite{BrigoCapponiPallaviciniPapatheodorou}, while, when we consider only margining costs, the formula is equal to the formula presented in the collateral section of this paper. In the following we consider some simple examples to highlight its meaning.

\subsubsection{Funding with Collateral}

If re-hypothecation is allowed, we assume that we can fund with collateral assets. Further, in order to obtain a more manageable set of equations, we assume that the hedging strategy is constituted by a set of rolling par-swaps, so that at each time the hedge portfolio's value process $H_t$ is zero. Thus the cash amount $F_t$ is given by
\[
F_t = {\bar V}_t(C;F) - C_t.
\]%
The choice for $F_t$ leads to a recursive equation which can be solved backward from maturity. Notice that the collateral account value $C_t$ is defined only at margining dates, but we are taking the limiting case of perfect collateralization, so that every time is a margining date (we recall that our definition of perfect collateralization requires that the mark-to-market of the portfolio is continuous). Further, being in the re-hypothecated case, we consider recoveries as given by $\rec_C'=\rec_C$ and $\rec_I'=\rec_I$.

If we assume also perfect collateralization, we get
\begin{eqnarray*}
{\bar V}_t(C;F) & = & \Ex{t}{\Pi(t,T\wedge\tau) + \gamma(t,T\wedge\tau;C) + \varphi(t,T\wedge\tau;F) } \\
                & + & \Ex{t}{\ind{\tau<T} D(t,\tau) C_\tau} \\
                & = & \Ex{t}{\Pi(t,T) + \gamma(t,T;C) + \varphi(t,T\wedge\tau;F)} \\
                & = & C_t + \Ex{t}{ \varphi(t,T\wedge\tau;F)}.
\end{eqnarray*}

Now, we consider two funding dates $t_j$ and $t_{j+1}$. By substituting the expression for funding cash flows we get (up to the default event)
\[
\left( {\bar V}_{t_j}(C;F) - C_{t_j} \right)^\pm = P^{f^\pm}_{t_j}(t_{j+1}) \left( \ExT{t_j}{t_{j+1}}{{\bar V}_{t_{j+1}}(C;F) - C_{t_{j+1}} } \right)^\pm,
\]%
where the expectation is taken under the $t_{j+1}$-forward measure. The terminal condition of the recursive equation can be calculated by noticing that
\[
\Ex{t}{ \varphi(t,T\wedge\tau;F)} |_{t=T\wedge\tau} = 0.
\]%
Hence, the BCCFVA price is equal to zero at each recursion step, leading to
\[
{\bar V}_t(C;F) = C_t = {\bar V}_t(C;0).
\]%
Thus, in case of perfect collateralization, there are no funding costs since we are {\em funding with collateral}.

\subsubsection{Funding without Collateral}

We assume that we are not collateralized. Thus, the cash amount $F_t$ needed for funding/investing is given by
\[
F_t = {\bar V}_t(0;F),
\]%
where we assume again that the hedge portfolio's value process $H_t$ is zero. Further, we consider a close-out amount inclusive of funding costs 
\[
\varepsilon_{I,\tau} \doteq \Ex{t}{\Pi(\tau,T) + \varphi(\tau,T;{\bar V}) }.
\]
Thus, we get
\begin{eqnarray*}
{\bar V}_t(0;F)
 & = & \varepsilon_{I,t} - \Ex{t}{\ind{\tau<T} \ind{\tau=\tau_C<\tau_I} D(t,\tau) (1-\rec_C)\varepsilon^+_{I,\tau} } \\
 & - & \Ex{t}{\ind{\tau<T} \ind{\tau=\tau_I<\tau_C} D(t,\tau) \left( (1-\rec_I)\varepsilon^-_{C,\tau} + (\varepsilon_{I,\tau}-\varepsilon_{C,\tau}) \right) }.
\end{eqnarray*}%

If we further specialize our example to the case of zero recovery, positive payoff, and negligible mismatch between close-out amounts calculated by the investor and the counterparty, we get
\[
{\bar V}_t(0;F) = \varepsilon_{I,t} - \Ex{t}{\ind{\tau<T} \ind{\tau=\tau_C<\tau_I} D(t,\tau) \,\varepsilon_{I,\tau} }.
\]%
Then, we consider two funding dates $t_j$ and $t_{j+1}$. By substituting the expression for funding cash flows we get (up to maturity)
\[
{\bar V}_{t_j}(0;F) = P^{f^+}_{t_j}(t_{j+1}) \ExT{t_j}{t_{j+1}}{ {\bar V}_{t_{j+1}}(0;F) + \ind{\tau=\tau_C>t_j} \frac{\Pi(t_j,t_{j+1})}{D(t_j,t_{j+1})} },
\]%
leading to
\[
{\bar V}_t(0;F) = \Ex{t}{ \sum_{j=1}^{m-1} \ind{\tau=\tau_C>t_j} \left( \prod_{i=1}^j \frac{P^{f^+}_{t_i}(t_{i+1})}{P_{t_i}(t_{i+1})} \right) \Pi(t_k,t_{k+1}) D(t,t_k) },
\]%
with $t_1=t$.

By taking the limit of continuous funding we get
\[
{\bar V}_t(0;F) = \Ex{t}{ \int_t^T \Pi(u,u+du) \, \exp\left\{ -\int_t^u dv\, \left( f^+_v + \lambda^{C<I}_v \right) \right\} },
\]%
where $\lambda^{C<I}_t$ is the counterparty's default intensity conditional on the investor not having defaulted earlier.

According to the liquidity policy in place, the funding rate $f^+_t$ may be specified by the treasury or, if the investor may directly fund on the market, it can be approximated by its bond/CDS spread $\ell^I_t$ plus risk-free rate $r_t$, as in \cite{MoriniPrampolini2011}, leading to
\[
{\bar V}_t(0;F) = \Ex{t}{ \int_t^T \Pi(u,u+du) \, \exp\left\{ -\int_t^u dv\, \left( r_v + \ell^+_v + \lambda^{C<I}_v \right) \right\} }.
\]%

\subsection{Recursive Solution of the BCCFVA Pricing Equation}

We derived in the previous sections the bilateral collateralized credit and funding valuation adjusted price equation \eqref{eq:bccfva} which allows to price a deal by taking into account counterparty risk, margining and funding costs. We built also some relevant examples to highlight the recursive nature of the equation and its link with discount curves. 

Now, we write a strategy to solve the equation without resorting to simplifying hypotheses. We try to explicit the recursion into an iterative set of equations which eventually are to be solved via least-square Monte Carlo techniques as in standard CVA calculations, see for instance \cite{BrigoPallavicini2007}.

We start by introducing the following quantities as building blocks for our iterative solution
\[
{\bar\Pi}_T(t_j,t_{j+1};C) := \Pi(t_j,t_{j+1}\wedge\tau) + \gamma(t_j,t_{j+1}\wedge\tau;C) + \ind{t_j<\tau<t_{j+1}} D(t_j,\tau) \theta_\tau(C,\varepsilon)
\]
where $\theta$ is still defined as in equation. (\ref{eq:theta}). The time parameter $T$ points out that the exposure $\varepsilon$ inside $\theta$ still refers to a deal with terminal maturity $T$. From the above definition it is clear that ${\bar\Pi}_T(t,T;C) = {\bar\Pi}(t,T;C)$. 

We solve equation \eqref{eq:bccfva} at each funding date $t_j$ in term of the price $\bar V$ calculated at the following funding time $t_{j+1}$, and we get
\begin{eqnarray*}
{\bar V}_{t_j}(C;F) &=& \Ex{t_j}{ {\bar V}_{t_{j+1}}(C;F) D(t_j,t_{j+1}) + {\bar\Pi}_T(t_j,t_{j+1};C) } \\
                    &-& \ind{\tau>t_j}
                          \left( F^-_{t_j} \frac{P_{t_j}(t_{j+1})}{P^{f^-}_{t_j}(t_{j+1})}
                               + F^+_{t_j} \frac{P_{t_j}(t_{j+1})}{{\bar P}^{f^+}_{t_j}(t_{j+1})}
                               - F_{t_j} \right).
\end{eqnarray*}

We recall that, if we assume $H_t=0$ as in the examples, we obtain that $F_{t_j} = {\bar V}_{t_j}(C;F)$ if re-hypothecation is forbidden, or $F_{t_j} = {\bar V}_{t_j}(C;F) - C_{t_j}$ if it is allowed. Furthermore, we have ${\bar V}_{t_n}(C;F) := 0$.

Hence, by solving for the positive and negative part when re-hypothecation is forbidden, we obtain, for $\tau<t_j$,
\[
\ind{\tau<t_j} {\bar V}_{t_j}(C;F) = \ind{\tau<t_j} \Ex{t_j}{ {\bar V}_{t_{j+1}}(C;F)D(t_j,t_{j+1}) + {\bar\Pi}_T(t_j,t_{j+1};C) },
\]%
while, for $\tau>t_j$,
\begin{equation}
\ind{\tau>t_j} {\bar V}^-_{t_j}(C;F) =
  \ind{\tau>t_j} P^{f^-}_{t_j}(t_{j+1})
  \left( \ExT{t_j}{t_{j+1}}{ {\bar V}_{t_{j+1}}(C;F) + \frac{{\bar\Pi}_T(t_j,t_{j+1};C)}{D(t_j,t_{j+1})} } \right)^-
\end{equation}%
and
\begin{equation}
\ind{\tau>t_j} {\bar V}^+_{t_j}(C;F) =
  \ind{\tau>t_j} {\bar P}^{f^+}_{t_j}(t_{j+1})
  \left( \ExT{t_j}{t_{j+1}}{ {\bar V}_{t_{j+1}}(C;F) + \frac{{\bar\Pi}_T(t_j,t_{j+1};C)}{D(t_j,t_{j+1})} } \right)^+.
\end{equation}%

On the other hand, when re-hypothecation is allowed, we obtain, for $\tau<t_j$,
\[
\ind{\tau<t_j} {\bar V}_{t_j}(C;F) = \ind{\tau<t_j} \Ex{t_j}{ {\bar V}_{t_{j+1}}(C;F)D(t_j,t_{j+1}) + {\bar\Pi}_T(t_j,t_{j+1};C) },
\]%
while, for $\tau>t_j$,
\begin{multline}
\ind{\tau>t_j} \left( {\bar V}_{t_j}(C;F) - C_{t_j} \right)^- = \\
  \ind{\tau>t_j} P^{f^-}_{t_j}(t_{j+1})
  \left( \ExT{t_j}{t_{j+1}}{ {\bar V}_{t_{j+1}}(C;F) + \frac{{\bar\Pi}_T(t_j,t_{j+1};C)-C_{t_j}}{D(t_j,t_{j+1})} } \right)^-
\end{multline}%
and
\begin{multline}
\ind{\tau>t_j} \left( {\bar V}_{t_j}(C;F) - C_{t_j} \right)^+ = \\
  \ind{\tau>t_j} {\bar P}^{f^+}_{t_j}(t_{j+1})
  \left( \ExT{t_j}{t_{j+1}}{ {\bar V}_{t_{j+1}}(C;F) + \frac{{\bar\Pi}_T(t_j,t_{j+1};C)-C_{t_j}}{D(t_j,t_{j+1})} } \right)^+.
\end{multline}%

\section{Conclusions: FVA and beyond}

We have seen above that when we try and include consistently funding, credit and collateral risk we obtain a highly nonlinear and recursive pricing equation, which is in line with recent findings in the literature, such as \cite{Crepey2011}. 

The first outcome from this analysis is that it is difficult to think of funding costs as an adjustment to be added to existing pricing equations, in the sense that something like
\begin{center}
{\em AdjustedPrice = RiskFreePrice + DVA - CVA + FVA}
\end{center}
cannot be obtained in a na\"ive way. Yet, one could define FVA formally as the difference between the price of the deal when all funding costs are set to zero and the price of the deal when funding costs are included, but such a definition will bring together both funding costs and CVA/DVA terms. In our notation
\[
\fva_t(C;F) = {\bar V}_{t}(C;0) - {\bar V}_{t}(C;F).
\]
The first term on the right hand side can be computed in a complex but non-recursive way as in \cite{BrigoCapponiPallaviciniPapatheodorou} or \cite{BrigoCapponiPallavicini}. The latter term needs to be computed by solving the recursive systems above through backward induction techniques.

Our message is that, similarly to how the CVA and DVA analysis made in \cite{BrigoCapponiPallaviciniPapatheodorou} shows that CVA and DVA cannot be considered simply as a spread on discounting curves, in the same way the present paper warns against considering FVA as a mere additive term of the pricing equation.

We argue that the exact nature of FVA can be seen from the examples we showed in the present paper. When proper conditions are set, we see that the recursive equation collapses into a modification of the pricing equation where the risk-free rate disappears. This is due to the fact that such conditions are exactly tuned so as to allow the investor to obtain all the cash he needs from an account at his disposal (the funding account $F_t$, or, in case of re-hypothecation, the collateral account $C_t$). In such situation the investor can fund or invest without resorting to a risk-free bank account (as he always does in practice).

Thus, we are tempted to make a further step and state that in realistic settings the pricing equations do not depend on the risk-free rate at all, but only on funding (and margining) rates. Yet, this step is to be taken with care. For instance, we should also start to include the funding costs of the counterparty in our framework, since on our default the close-out amount is calculated by her. We leave such analysis to further work

\newpage

\bibliographystyle{plainnat}
\bibliography{funding}

\end{document}